\documentclass[aps,pra,twocolumn,superscriptaddress,showpacs]{revtex4-1}
\usepackage[utf8]{inputenc}

\usepackage{graphicx}

\graphicspath{{./fig/}{./}}

\usepackage{hyperref}
\usepackage{amsmath}
\usepackage{amssymb}
\usepackage{epstopdf}
\usepackage{color}

\newcommand{\red}[1]{{\color{red}#1}}

\newcommand{\ts}[1]{}

\newcommand{\bra}[1]{\langle #1 \vert} 
\newcommand{\ket}[1]{{\vert#1\rangle}} 
\newcommand{\abs}[1]{\left|#1\right|} 
\newcommand{\mean}[2] {  \langle  #1 \rangle _{#2} }

\newcommand{\citel}[1][]{%
	\red{\ifthenelse{\equal{#1}{}}{[?]}{[#1]}}%
}
\renewcommand{\Im}{\operatorname{Im}}
\renewcommand{\Re}{\operatorname{Re}}

\newcommand{\Nwf}[0] {N_\text{wf}}
\newcommand{\Nmode}{N_\text{mode}}
\newcommand{\Nsamp}{N_\text{samp}}

\newcommand{\skp}[1]{}

\newcommand{\Fig}{Fig.\@ }
\newcommand{\Eq}{Eq.\@ }
\newcommand{\Ref}{Ref.\@ }

\begin{document}

\title{Accurate photonic temporal mode analysis with reduced resources}

\author{O. Morin, S. Langenfeld, M. Körber, and G. Rempe }
\affiliation{Max-Planck-Institut f\"{u}r Quantenoptik, Hans-Kopfermann-Strasse 1, 85748 Garching, Germany}
\date{\today}

\begin{abstract}
The knowledge and thus characterization of the temporal modes of quantum light fields is important in many areas of quantum physics ranging from experimental setup diagnosis to fundamental-physics investigations. Recent results showed how the auto-correlation function computed from continuous-wave homodyne measurements can be a powerful way to access the temporal mode structure. Here, we push forward this method by providing a deeper understanding and by showing how to extract the amplitude and phase of the temporal mode function with reduced experimental resources. Moreover, a quantitative analysis allows us to identify a regime of parameters where the method provides a trustworthy reconstruction, which we illustrate experimentally.
\end{abstract}


\maketitle

Techniques to characterize quantum states have become more and more valuable with the development of quantum information science. In order to establish standards and check the compliance of the different building blocks, it is necessary to have efficient and reliable measurement methods. However, characterizing quantum states is challenging by nature \cite{Wootters1982}. One quantum measurement only provides a limited amount of information and therefore it is always necessary to perform multiple measurements to obtain a full quantum state characterization. All the art lies in the way of assembling those pieces of information \cite{Leonhardt, Helstrom}. 

Beyond the conceptual interest, it is also important to pay attention to the practical aspects: Each measurement demands experimental resources, e.g., time, energy, money and processing power. Regarding the valuable character of the measurements, it is thus essential to know how the choice of the experimental parameters, on one hand, and the evaluation technique, on the other hand, can maximize the amount of extracted information. In other words, the question is how to obtain the desired characterization at reduced \emph{costs}.

Optical states constitute the essential ingredient of many quantum information protocols, for quantum-network applications \cite{Kimble2008, Wehner2018} as well as all-optical processing \cite{Minzioni2019}. Among the different degrees of freedom that define an optical mode, the temporal shape has lately arisen an increasing interest \cite{Brecht2016}. However, it appears to be challenging to control \cite{Raymer}. Although theoretical models have been developed for various systems \cite{Goschkov2007,Nielsen2007}, they don't always accurately describe the experimental implementation. Hence, in order to verify, experimental techniques to measure the temporal mode are necessary.  The Hong-Ou-Mandel experiment was probably the first technique looking closely at the characterization problem of temporal modes \cite{HOM}. Although it gives information about the coherence of the temporal shape, this technique evaluates the degree of indistinguishability and does not provide any information about the details of a possible mismatch. In addition, it remains restricted to single-photon states. 

Recently new methods have been developed to tackle the challenge in practical cases \cite{Bellini,MorinMode,Lvovsky,Huang,Baune2017,Yang2018,Takase}. Those techniques are mostly based on homodyne detection. 
With some analogies with the methods developed for the spatial degree of freedom \cite{Dawes2003}, recent techniques have shown that the use of the auto-correlation function from homodyne measurements is an efficient strategy. However, in the technique presented in \Ref \cite{MorinMode}, only real-valued temporal mode functions can be reconstructed. Reference \cite{Lvovsky} proposed a solution to reconstruct temporal mode functions with complex values by performing measurements at various detunings of the local oscillator. Those additional measurements require a sophisticated algorithm in order to extract the temporal density matrix. Another procedure has been proposed in \cite{Takase}. There, one frequency is sufficient but then the use of a double homodyne detection is required. 

\begin{figure}[!t]
\includegraphics[width=0.95\columnwidth]{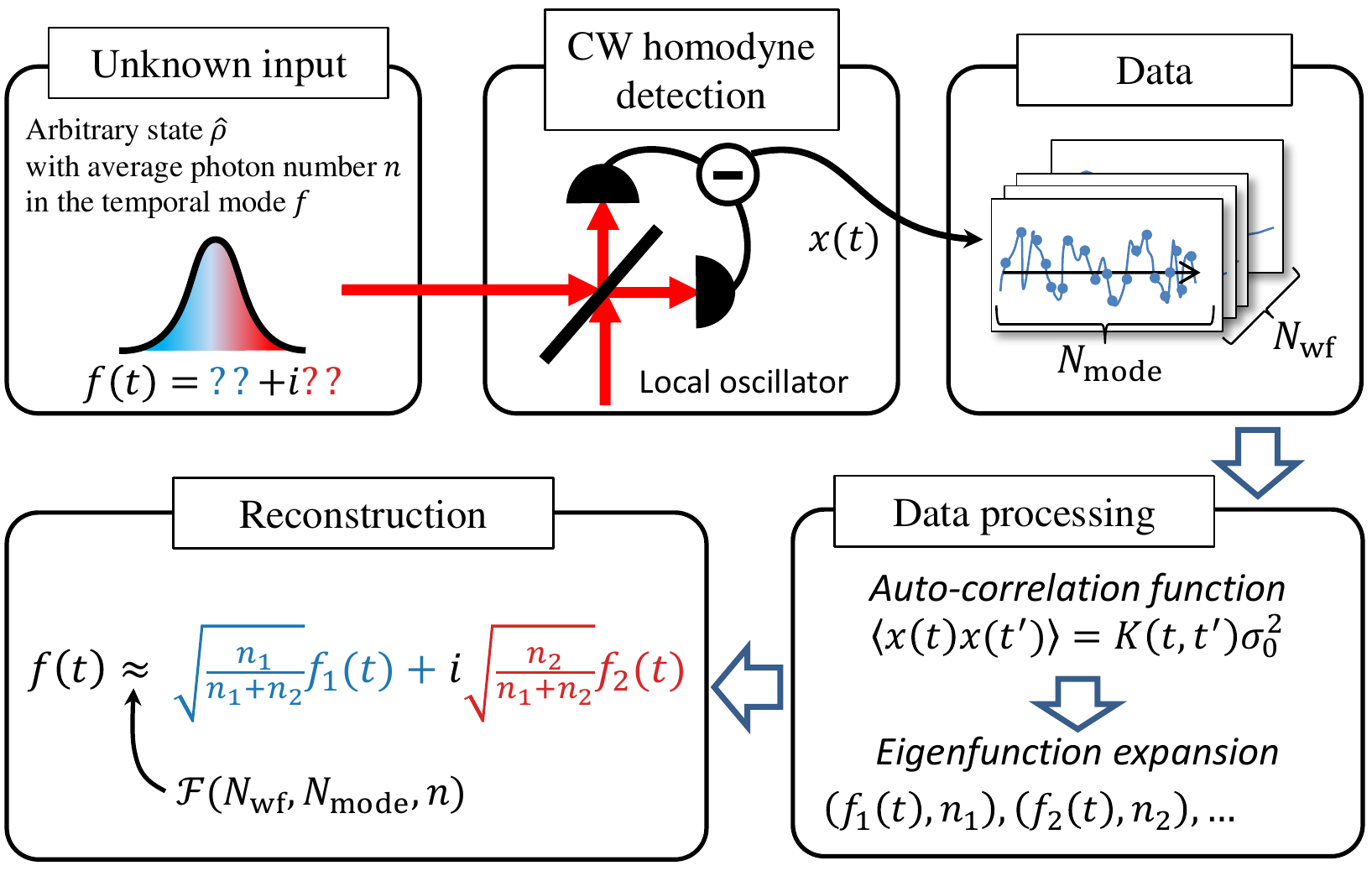}
\caption{(color online) Principle of the temporal  mode reconstruction. The state with an unknown complex-valued temporal shape $f$ is measured by a cw homodyne detection. The collected data are used to compute the auto-correlation function. Then, by applying the eigenfunction expansion, we use two of the eigenfunctions and associated eigenvalues to reconstruct the temporal mode. Knowing the average photon number $n$, the accuracy of the reconstruction can be set by the choice of the number of waveforms $\Nwf$ and the number of measured modes $\Nmode$.}
\label{fig:principle}
\end{figure}

Following those works, here we provide new insights regarding the problem of experimental costs we just underlined. Starting from reasonable assumptions, we propose an alternative to extract the complex-valued temporal mode function without the need of additional measurements nor increasing the complexity of the data processing or experimental apparatus. In addition, we also address the question of the accuracy of the results and evaluate the conditions under which a maximum of valuable information can be extracted. Therefore, we provide here a turnkey method that can be successfully used in a wide range of systems beyond the usual continuous-variables framework.

\emph{Temporal mode reconstruction.} The reconstruction procedure is depicted in \Fig \ref{fig:principle}. The signal recorded by a continuous wave (cw) homodyne detection is a time dependent measurement of the quadrature observable $\hat{x}_\theta(t)$. Here, $\theta$ represents the phase between the local oscillator and the measured state. In the following, we assume that all the recordings are done with phase averaging (see Supplemental Material (SM)) and thus omit the notation in the following. 
From those quadrature signals we compute the auto-correlation function
$\mean{\hat{x}(t)\hat{x}(t')}{}$ which can be represented as $K(t,t')\cdot\sigma_0^2$. The associated kernel $K$ is independent of the variance of the vacuum $\sigma_0^2=\bra{0}\hat{x}^2\ket{0}$.

Assuming the measurement of an arbitrary state in a pure single temporal mode defined by the function $f$, e.g.\@ a single-photon state $\hat{a}^\dagger_f\ket{0}=\int dt f(t) \hat{a}^\dagger\ket{0}$, the kernel is of the form
\begin{equation}
K(t,t')=\delta(t-t')+2\mean{\hat{n}_f}{}\Re[f(t)f^*(t')]\ ,
\label{eq:start_eq}
\end{equation}
where $\mean{\hat{n}_f}{}$ is the average photon number (SM). 
As shown in \cite{MorinMode}, if $f$ is real, one can immediately identify that it is also an eigenfunction of the kernel $K$ with the associated eigenvalue $\kappa=2n_f+1$. Any other function orthogonal to $f$ is an eigenfunction too, but with an eigenvalue equal to 1. Hence, by measuring $K$, the function $f$ will correspond to the eigenfunction associated with the largest eigenvalue, i.e., $\kappa>1$.

If $f$ is a function with complex values, we have  to consider more than one eigenfunction. Indeed, the second term of Eq. (\ref{eq:start_eq}) can be  expanded in the following way
\begin{equation}
\Re[f(t)f^*(t')]=\Re[f(t)]\Re[f(t')]+\Im[f(t)]\Im[f(t')]\ .
\end{equation}
Hence, we recognize the eigenfunction expansion which is always guaranteed by Mercer's theorem \cite{Courant} since $K(t,t')$ is symmetric positive. If we call $\phi$ the argument of the function $f$ such that $f(t)=\abs{f(t)}e^{i\phi(t)}$, we have the two eigenfunctions of $K$
\begin{align}
f_1(t) &=\sqrt{\tfrac{n_1+n_2}{n_1}}\abs{f(t)}\cos(\phi(t)+\phi_0)\ ,\\
f_2(t)&=\sqrt{\tfrac{n_1+n_2}{n_2}}\abs{f(t)}\sin(\phi(t)+\phi_0)\ .
\end{align}
Those two eigenfunctions are associated with the eigenvalues $\kappa_i = 2n_i+1$ with $n_i$ being the average photon number in the mode $f_i$. The energy conservation is satisfied by $ \mean{\hat{n}_f}{} = n_1+n_2$. Note that $\phi$ can be redefined up to any arbitrary global phase, and here, $\phi_0$ only ensures the orthogonality for a given $\phi(t)$,  i.e., $\int f_1(t)f_2(t) dt=0$. As for the case of $f$ being real, all the other eigenfunctions, being orthogonal to $f_1$ and $f_2$, are associated to the eigenvalues equal to 1, thus corresponding to the vacuum states. 
Eventually, the temporal mode function is reconstructed with the two eigenfunctions and their associated average photon numbers
\begin{equation}
f(t)=\frac{1}{\sqrt{n_1+n_2}}\left(\sqrt{n_1}f_1(t)+i\sqrt{n_2}f_2(t)\right) \ .
\label{eq:complet_mode}
\end{equation}

However, with only one frequency for the local oscillator, an ambiguity on the sign of the phase remains. Indeed, without prior knowledge about $f$, there is no way to discriminate which eigenfunction is $f_1$ and which one is $f_2$. In other words, which eigenfunction has to be taken as the real part and which one is the imaginary part of $f$.
This is analog to the case of the beating signal of two waves: Only the absolute difference of frequencies is measured and it is not possible to tell which one of the two waves has the larger frequency. However, as soon as one has a second reference, this ambiguity can be lifted. Eventually, in practice, it is not unusual that one has some knowledge about the characterized photonic state and, depending on the tested physical system, one can maybe infer information about which sign makes sense. 

\emph{Multimode and statistical mixture.} As mentioned previously, to apply the proposed reconstruction we need to assume that the measured state lies in a pure single temporal mode. (Note that here the purity concerns the mode but that the state within the mode can be a statistical mixture though.) However, this assumption can be checked \emph{a posteriori}.

Three cases are possible. First, if only one eigenvalue is greater than one, there is no ambiguity, the state is in a pure real-valued temporal mode. Second, if more than two eigenvalues are greater than one, then there are more than two modes involved. What can be said about the details of these modes is beyond the scope of this paper and probably requires some assumptions. For example, in Ref. \cite{MorinMode, Huang} all temporal modes are assumed to be pure and with real values. Third, if two eigenvalues are above one, either there is a pure complex-valued temporal mode, or there are two modes (mixed or pure). Nevertheless, the pure single-mode case can be confirmed by performing a new reconstruction with the compensating phase $\phi(t)$ on the local oscillator, effectively making $f$ real-valued. This way, the new reconstruction should exhibit only one eigenvalue above one. If not, then we can conclude that the mode was not pure and/or single.

\emph{Reconstruction accuracy.} For any reconstruction method it is important to be able to evaluate its accuracy. For non-trivial cases like here, the usual approach consists in performing multiple simulations: Given some results, one can estimate the uncertainty and therefore how close these results are from the real state. However, the accuracy is known only \emph{a posteriori}. Considering the \emph{experimental costs} of the measurements, it can be valuable to know in advance how accurate the reconstruction can be. 

Practically, the implementation of the method requires some unavoidable parameter choices. In our reconstruction method, three parameters have to be considered:
\begin{itemize}
\item $\Nwf$, the number of waveforms acquired from the homodyne signal,
\item $\Nmode$, the number of modes which are measured,
\item $\mean{\hat{n}_i}{}$, the average photon number per mode $i$.
\end{itemize}

First, the number of measured waveforms used to compute the kernel $K(t,t')$ is finite which will obviously lead to some statistical limitations.

Second, each waveform will be recorded at a finite rate and for a finite duration. Ideally, the number of samples per waveform $\Nsamp$ simply corresponds to the number of modes. If we try to get a description closer to the experimental implementation, one has to consider the different electronic stages from the photodiodes to the analog-to-digital converter which define an overall limited bandwidth. Therefore, the different modes are not measured with the same gain. Hence, the effective number of measured modes is in this case given by $\Nmode \propto \Delta \omega/\delta \omega$, namely the ratio of the bandwidth by the resolution bandwidth. The case $\Nmode=\Nsamp$ is then only true if the bandwidth is larger than the sampling frequency $1/\delta t$. 

Eventually, the average photon number $\mean{\hat{n}}{}$ is not strictly speaking an experimental parameter as it cannot be adjusted. Nevertheless, improving the efficiency of a homodyne detection can be challenging and therefore it is a valuable information to know how rewarding the improvement of the efficiency of the setup would be.

\begin{figure}[!t]
\includegraphics[width=0.99\columnwidth]{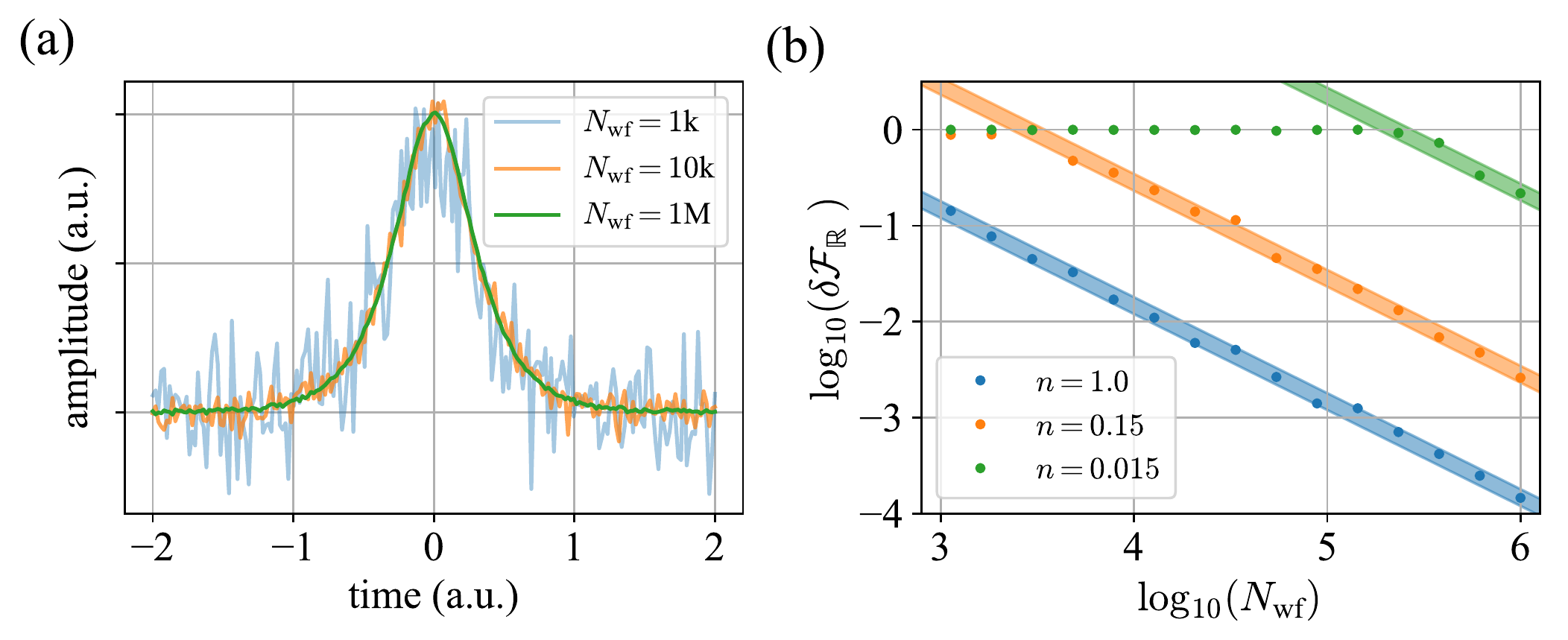}
\caption{(color online) (a) Temporal shape for different values of $\Nwf$. (b) Infidelity versus $\Nwf$ for different values of the average photon number $n$. The colored areas correspond to the 2$\sigma$ distributions, i.e., $\mu(\delta\mathcal{F})\pm2\sigma(\delta\mathcal{F})$. For (a) and (b) $\Nmode=\Nsamp=200$.}
\label{fig:eigenmode}
\end{figure}

In order to quantify the impacts of those parameters on the results, we have performed multiple Monte-Carlo simulations and compared input $(f,n)$ and output $(f_\text{meas},n_\text{meas})$. From this we infer the dependency between the results accuracy and the different parameters. We do not make any mathematical proof and therefore the conclusion we provide should be considered as a help for performing measurements. Most of the raw simulation results are reported in the SM.

The similarity between $f$ and $f_\text{meas}$ is usually quantified by the fidelity $\mathcal{F}=\abs{\int dt f^*(t)f_\text{meas}(t)}^2$.
Here, as the fidelity will be ideally close to unity, we analyze the infidelity $\delta\mathcal{F}=1-\mathcal{F}$. Concerning the evaluation of the photon number, we look at the deviation defined by $\delta n= n_\text{meas}-n$. As shown by \Eq (\ref{eq:complet_mode}), when $f$ has complex values, the reconstruction requires to combine two eigenmodes. For simplicity, we first focus on the reconstruction involving one eigenmode. This corresponds to the case of $f$ being real, in which case we label the fidelity $\mathcal{F}_\mathbb{R}$. This case also applies to complex-valued functions when looking at the real/imaginary part individually.

Hence, over multiple simulations, we have found that the average value of the infidelity has the form
\begin{equation}
\mu(\delta\mathcal{F}_\mathbb{R})\approx\frac{\Nmode/2}{\Nwf}\cdot \frac{1}{n}\left(1+\frac{1}{2n}\right)
\label{eq:av_fid}
\end{equation}
and a standard deviation
\begin{equation}
\sigma(\delta\mathcal{F}_\mathbb{R})\approx\frac{\sqrt{\Nmode/2}}{\Nwf}\cdot\frac{1}{n}\left(1+\frac{1}{2n}\right)\  .
\label{eq:std_fid}
\end{equation}

Figure \ref{fig:eigenmode}(a) shows the outcome of the temporal mode function reconstruction $f_\text{meas}$ for the simulation of a real function $f$ (with $n=1$ and $\Nmode=\Nsamp=200$). We can see that even with a small number of waveforms the shape can already be observed. However, the noise decreases with increasing  $\Nwf$. 

Interestingly, \Fig \ref{fig:eigenmode}(b) shows that there is a clear threshold when $\mu(\delta\mathcal{F})=1$ which basically means that no $f_\text{meas}$ can be reconstructed or in other words, $f_\text{meas}$ has no overlap with $f$. Hence, with an \emph{a priori} coarse knowledge of $n$, the choice of the parameters should fulfill the condition $\sqrt{\frac{\Nmode}{\Nwf}}\ll n$.

Another important message here is that a too large number of modes is detrimental for the accuracy. This is actually counter-intuitive as it means that an excessively large electronic bandwidth (or a too high time resolution) and/or a too small frequency resolution (or too long waveform duration) reduces the achievable accuracy.

Conversely, $\Nmode$ cannot be too small either as with an under-sampling the details of the shape would be missing. 
Here, this is not caught by the simulations. Indeed, a continuous function and a discretized function don't have a unity overlap but here we compute the fidelity between $f$ and $f_\text{meas}$ both being sampled the same way. 
Although that is a trivial problem, it is really specific to each temporal mode shape and therefore not possible to study in the most general case. 
In other words, in relation to the Nyquist-Shannon theorem, this defines a lower bound on $\Nmode$.

\begin{figure}[!t]
\includegraphics[width=0.99\columnwidth]{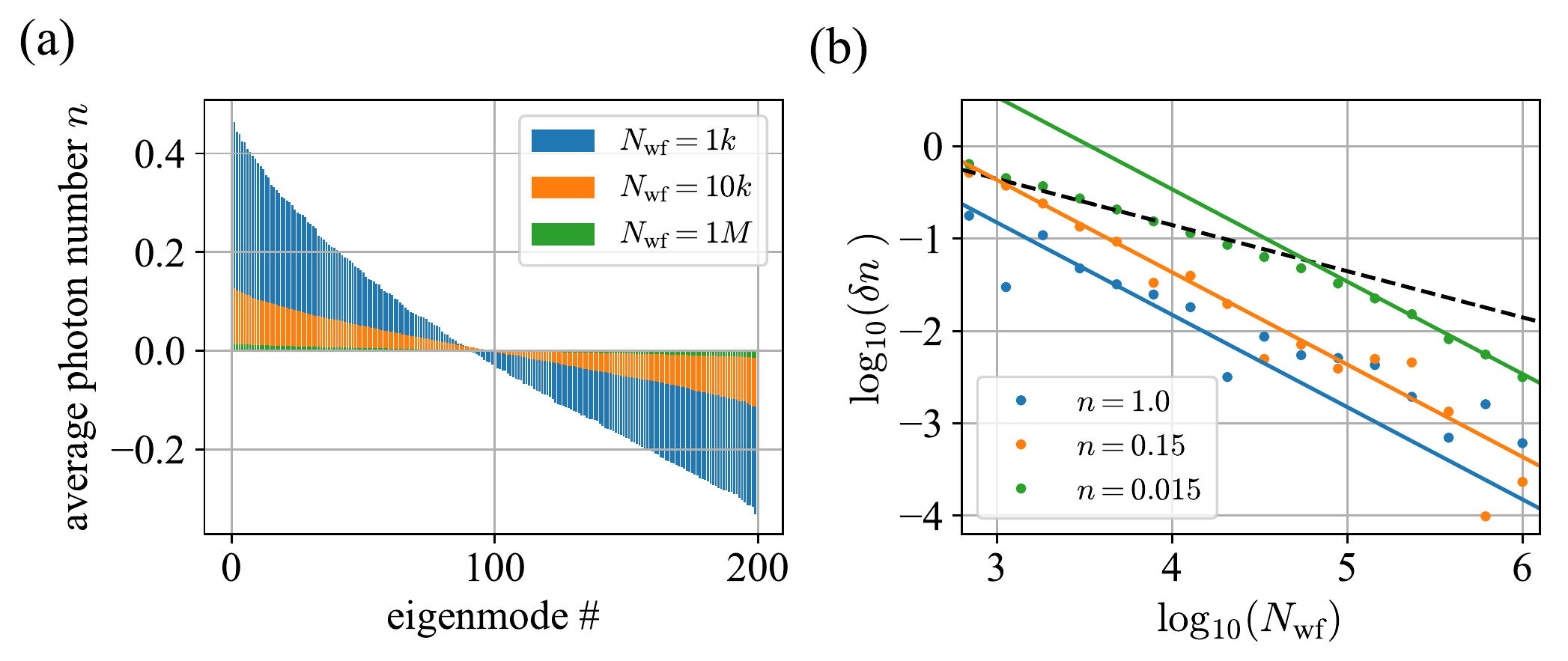}
\caption{(color online) (a) Average photon number of vacuum states obtained from the eigenfunction expansion for different $\Nwf$. Ideally all values should be zero but the statistical limitation given by $\Nmode$ and $\Nwf$ leads to a deviation. (b) Scan of $\Nwf$ for different values of $n$. The plain line gives $\mu(\delta n)$ from equation (\ref{eq:mudn}). The black line is the upper bound given by the vacuum state $\mu(\delta n)_{\ket{0}}\approx\sqrt{\Nmode/\Nwf} $. }
\label{fig:eigenvalue}
\end{figure}

As shown by \Eq \ref{eq:complet_mode}, the average photon numbers obtained from the eigenvalues are part of the reconstruction when considering a complex-valued temporal mode function. Therefore, it is important to understand their accuracy too. In \Fig \ref{fig:eigenvalue}(a), for a vacuum state on every mode, the average photon number for each mode can be above or below zero. This is due to the statistical limitation (but not exclusively as we will see next). Hence, by increasing the number of waveforms $\Nwf$, we see that this deviation decreases. We find that on average the deviation follows
\begin{equation}
\mu(\delta n)_{\ket{0}}\approx\sqrt{\frac{\Nmode}{\Nwf}}\ .
\end{equation}
In \Fig \ref{fig:eigenvalue}(b) we can see that this formula (dashed black line) is also the bound  for states different than vacuum. Beyond this bound, when $n\neq0$,  the average value of the deviation $\delta n$ is
\begin{equation}
\mu(\delta n)\approx\frac{\Nmode/2}{\Nwf}\left(1+\frac{1}{2n}\right) \ .
\label{eq:mudn}
\end{equation} 
Therefore, one can only measure modes with photon numbers $n$ by fulfilling the condition $n \gg \sqrt{\frac{\Nmode}{\Nwf}}$ which is similar to the condition found for the fidelity of the temporal function. This bound is well illustrated in \Fig \ref{fig:eigenvalue}(b) by the case of $n=0.015$.

The standard deviation for $n$ is not specific to the reconstruction we use here. It is limited by the amount of data even when knowing the temporal mode $f$ or using single photon counting. In addition, it is specific to the photon number distribution of the quantum state. Thus we cannot provide any general formula.

As announced earlier, all the previous results apply to the case of single eigenmodes. Now, knowing how accurate each individual mode is, we are interested in the full reconstruction given by \Eq (\ref{eq:complet_mode}) which involved two eigenmodes. Due to this combination of eigenmodes, the analysis of the accuracy is less straightforward. Indeed, the division of the global average photon number $n$ between $n_1$ and $n_2$ is not independent of the temporal mode function $f$. However, we can distinguish two extreme cases: First, when the temporal mode function is real and second, when it is complex with balanced contributions on the real and imaginary parts.

When the temporal mode function $f$ is real but no \textit{a priori} assumption can be made, the contribution of the second mode necessarily degrades the fidelity. This contribution essentially comes from the non-zero photon number measured for the second mode. As shown earlier, even the vacuum state shows a deviation in the photon number $\mu(\delta n)_\ket{0}$. Hence, the fidelity will be bounded by  $\mu(\delta n)_\ket{0}/n$. In this case the scaling is less favorable compared to the reconstruction when assuming a real function. 
A possible strategy to improve the accuracy would be to split the dataset into two in order to minimize the error on $n_2$.

The second extreme case is when the complex-valued function $f$ leads to a balanced division of the average photon number, i.e., $n_1=n_2$. In that case, the average fidelity is straightforward to calculate and is equal to the one in \Eq (\ref{eq:av_fid}) but the average photon number has to be replaced by $n/2$.

In summary, without any assumption on $f$ and more precisely on the photon number division $n_1$ and $n_2$, the fidelity of the reconstructed state will be bounded by those two aforementioned cases
\begin{equation}
\frac{\Nmode/2}{\Nwf} \frac{1}{n/2}\left(1+\frac{1}{n}\right) \lesssim\mu(\delta \mathcal{F}_\mathbb{C})\lesssim \frac{1}{n} \sqrt{\frac{\Nmode}{\Nwf}} \ .
\label{eq:boundaries}
\end{equation}

\begin{figure}[!t]
\includegraphics[width=0.99\columnwidth]{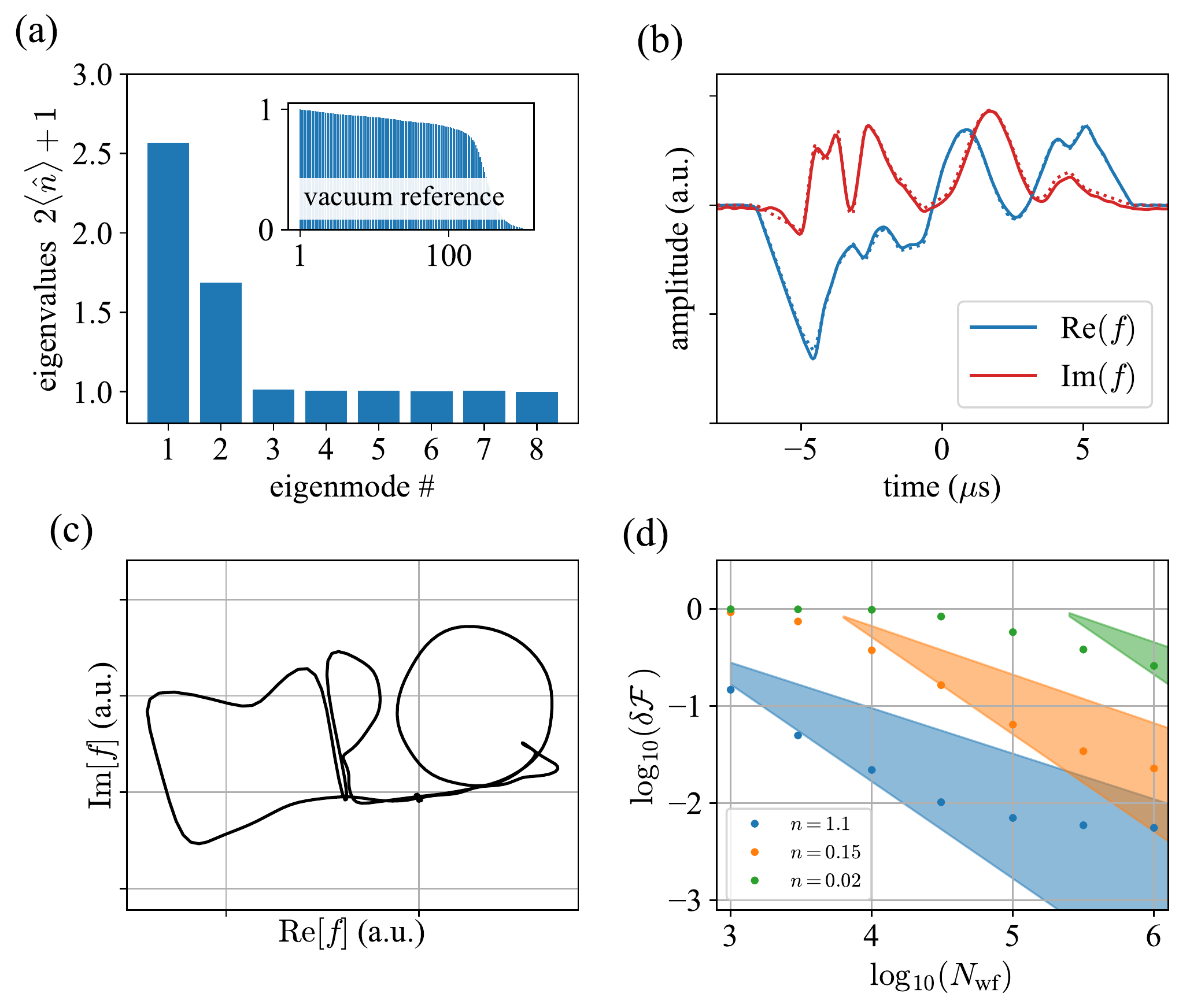}
\caption{(color online) Experimental demonstration. (a) Spectrum of eigenvalues. The inset is the spectrum measured for the vacuum state input that is used to determine the relevant number of modes measured by the homodyne detection. (b) Eigenfunctions that correspond to the real and imaginary part of the temporal mode functions. Plain curves are the measured functions and dashed the targeted ones. (c) Polar plot of the temporal mode function reconstructed via \Eq (\ref{eq:complet_mode}). The fidelity with the target shape ``MPQ" is equal to $\mathcal{F}=99.44\%$. This dataset contains $\Nwf = 1\text{M}$ waveforms and the input probe state an average photon number of $n=1.1$. (d) Infidelity as a function of the number of waveforms $\Nwf$ for three values of $n$. Regarding the inset of (a), we consider $\Nmode=100$. The colored areas correspond to the limits defined by \Eq (\ref{eq:boundaries}).
}
\label{fig:exp}
\end{figure}

\emph{Experimental illustration.} In the following we check our method experimentally. As mentioned in the introduction, we restrict our study to the impact of statistical limitations. 
The effect of experimental imperfections is a relevant topic too, but cannot be modeled in a general way. However, we assume that our system is good enough and we show how the various conclusions we have drawn are applicable and thus fruitful practically.

We illustrate our method with a temporal mode with the shape ``MPQ'' in the polar representation. This shape has the advantage to probe the technique with a relatively sophisticated shape in amplitude and phase. The duration and dynamics of the shape are chosen to be within the band of the homodyne detection (band-pass of 10kHz-3MHz). The probe state is a coherent state obtained via a weak cw laser beam sent through an acousto-optic modulator such that the desired shape is imprinted on the beam.

Figure \ref{fig:exp} shows the experimental results. On the histogram (a) we see the two relevant eigenvalues for the reconstruction. We also show in the inset the spectrum of the vacuum state. This later allows us to estimate the number of modes measured by the homodyne detection and therefore estimate the accuracy of the reconstruction. The plot (b) depicts the two eigenfunctions associated with the two eigenvalues significantly above 1. The plain lines represent the measured functions whereas the dotted lines represent the theoretical ones. In \Fig (c) we plot those two functions in a parametric way which corresponds to the polar representation of the complex-valued temporal mode.

Eventually, we check how the fidelity changes by repeating the same reconstruction for different average photon numbers and various number of waveforms. In plot (d), we show $\delta\mathcal{F}$ as a function of the number of waveforms $\Nwf$. The two bounds described in \Eq (\ref{eq:boundaries}), namely the balanced and unbalanced mode distributions, give a satisfying range of accuracy for the reconstructed temporal functions. One can see, however, that the fidelity settles for high $\Nwf$. We attribute this to the experimental precision of the generation of the probe state, one possible issue being the calibration of the non-linear response of the acousto-optic modulator. Furthermore, we cannot exclude, regardless of the source, some phase and/or amplitude noise. This latter issue should translate into more eigenvalues above $1$ but at such a low level that it is difficult to measure those eigenvalues.

\emph{Conclusion.} Here, we have provided a full recipe to reliably characterize any arbitrary complex-valued temporal mode function. This method features an extremely simple implementation in terms of the experimental setup as well as data processing. Importantly, we provide all the key aspects that are mandatory to guarantee a trustworthy reconstruction with a reduced amount of resources. This can be a game changer as we have shown that some aspects can be counter-intuitive at first sight. 
In addition, we have illustrated the method on an experimental case and confirmed that it is a powerful way to reconstruct even a sophisticated shape. 
Hence, we believe that this will facilitate the implementation of the presented technique beyond the continuous variables community and make this technique a standard routine in quantum optics laboratories \cite{MorinPRL2019}.

\begin{acknowledgments}

The authors thank S. Dürr for fruitful discussions. This work was supported by the Bundesministerium für Bildung und Forschung via the Verbund Q.Link.X (16KIS0870), by the Deutsche Forschungsgemeinschaft under Germany’s Excellence Strategy – EXC-2111 – 390814868, and by the European Union’s Horizon 2020 research and innovation programme via the project Quantum Internet Alliance (QIA, GA No. 820445).

\end{acknowledgments}

\appendix

\section{Generalization to any state in a single mode}

We show in the following one possible way to derive \Eq (1) of the main text. (The proof is relatively straightforward in the case of a single-photon state). The auto-correlation function of the quadrature operators can be expressed as a function of the annihilation and creation operators
\begin{equation}
\begin{split}
\hat{x}_\theta(t)\hat{x}_\theta(t')=&\left(e^{-i\theta}\hat{a}(t)+e^{i\theta}\hat{a}^\dagger(t)\right)\left(e^{-i\theta}\hat{a}(t')+e^{i\theta}\hat{a}^\dagger(t')\right)\\
=&\ e^{-2i\theta}\hat{a}(t)\hat{a}(t')+e^{2i\theta}\hat{a}^\dagger(t)\hat{a}^\dagger(t')\\
&+\left(\hat{a}(t)\hat{a}^\dagger(t')+\hat{a}^\dagger(t)\hat{a}(t')\right) \ .
\end{split}
\end{equation}
First, due to the phase averaging, 
\begin{equation}
\hat{x}(t)\hat{x}(t')=\frac{1}{2\pi}\int_{[0,2\pi]}\hat{x}_\theta(t)\hat{x}_\theta(t')\ d\theta  \ ,
\end{equation}
the first two terms vanish. Second, by using the commutation relation $[\hat{a}(t),\hat{a}^\dagger(t')]=\delta(t-t')$ we obtain
\begin{equation}
\hat{x}(t)\hat{x}(t')= \delta(t-t')+\hat{a}^\dagger(t')\hat{a}(t)+\hat{a}^\dagger(t)\hat{a}(t') \ .
\label{eq:product}
\end{equation}
We call $f$ the temporal mode in which the measured state $\hat{\rho}$ lies. Hence, we can define a set of orthonormal functions $\{f_k\}$ with $f=f_{k_0}$. With this set of functions forming a complete basis, we have $\hat{a}(t)=\sum_k f_k(t)\hat{a}_k$. Hence, we can write
\begin{equation}
\hat{a}^\dagger(t)\hat{a}(t')=\sum_k\sum_{k'} f_k^*(t)f_{k'}(t')\hat{a}_k^\dagger\hat{a}_{k'} \ .
\end{equation}
As we assume a single-mode state, all other temporal modes $f_{k\neq k_0}$ contain the vacuum state. Therefore, we have $\mean{\hat{a}^\dagger_{k_0}\hat{a}_{k_0}}{}=\mean{\hat{n}_f}{}$ and, for any $k\neq k_0$ or $k'\neq k_0$, we have  $\mean{\hat{a}^\dagger_{k}\hat{a}_{k'}}{}=0$. The average of Eq. (\ref{eq:product}) becomes
\begin{align}
\mean{\hat{x}(t)\hat{x}(t')}{}&=\delta(t-t')+\mean{\hat{n}_f}{}\left[f^*(t)f(t')+f^*(t')f(t)\right] \ ,\\
&=\delta(t-t')+2\mean{\hat{n}_f}{}\Re[f^*(t)f(t')] \ .
\end{align}

It is important to note that no assumptions are made on the input state beside the fact that it lies in a pure single temporal mode. This proves that the reconstruction of the temporal mode function is independent of the state $\hat{\rho}$ and that only the average photon number plays a role in the reconstruction. One can measure the temporal mode function regardless of the state within that mode (Fock states, coherent states, squeezed vacuum, ...).

\section{Homodyne signal simulation}

The signal $x(t)$ is actually discrete such that $x(t_i)=x_i$ and limited in duration such that $i\in [0,\Nmode-1]$.
For each homodyne signal and for each mode $f_k$ we need to generate $\Nmode$ random values $x_k$. 
In the case of a temporal mode function with real values only, only one mode can potentially contain one photon, all the others being the vacuum state. Thus, we need one random number $x_0$ following the single photon distribution $\mathcal{P}_\ket{1}(x)$ and all the others $x_{k\neq 0}$ following the vacuum distribution $ \mathcal{P}_\ket{0}$ (Gaussian distribution). From this set, and by defining $f_0$ the mode of the single photon state, one can compute the homodyne signal 
\begin{equation}
x(t_k)=\sum_{j=1}^{\Nmode}x_j f_j(t_k)
\end{equation}
It is then possible to apply at this stage low- and/or high-pass filters. However, it is safer to then start with a longer time segment and truncate it after the filtering in order to minimize edge effects.

The case of a temporal mode with complex values is more tricky. First one needs to split the temporal function into real and imaginary part. One obtains
\begin{align}
\int f(t)\hat{a}^\dagger(t)dt\ket{0}&=\int[tf_r(t)+irf_i(t)]\hat{a}^\dagger(t)dt\ket{0}\\
&=[t\hat{a}_r^\dagger+ir\hat{a}_i^\dagger]\ket{0}
\end{align}
where $r^2+t^2=1$.
We thus have a conditional probability between the two modes such that the probability to measure $x_r$ in the real mode and $x_i$ in the imaginary mode is
\begin{equation}
\begin{split}
\mathcal{P}_{r,i}(x_r,x_i)=\ &\eta\abs{\bra{x_r}\bra{x_i}[t\hat{a}_r^\dagger+r\hat{a}_i^\dagger]\ket{0}}^2\\
&+(1-\eta)\abs{\bra{x_r}\bra{x_i}\ket{0}}^2\\
=\ &\eta\left(t^2\mathcal{P}_\ket{1}(x_r)\mathcal{P}_\ket{0}(x_i)+r^2\mathcal{P}_\ket{0}(x_r)\mathcal{P}_\ket{1}(x_i)\right)\\
&+(1-\eta)\mathcal{P}_\ket{0}(x_r)\mathcal{P}_\ket{0}(x_i).
\end{split}
\end{equation}
Practically, this results in generating $x_r$ with the distribution
\begin{equation}
\mathcal{P}_r(x_r)=\eta\left(t^2\mathcal{P}_\ket{1}(x_r)+r^2\mathcal{P}_\ket{0}(x_r)\right)+(1-\eta)\mathcal{P}_\ket{0}(x_r)
\end{equation}
and then, for the given $x_r$, one generates $x_i$ following the distribution $\mathcal{P}_{r,i}(x_r,x_i)$.

Regarding the temporal mode analysis process, the correlation is in principle unimportant and random numbers could be chosen independently for the two modes. However, it would be questionable to apply this assumption here as we want to check that the procedure is giving the expected results.

We have also performed simulations for coherent states. These simulations do not present this problem of correlated modes. Indeed, by essence a coherent state doesn't lead to entanglement via linear operations. On the other hand, in contrast to the single photon state, a coherent state has a phase. Hence, in that case one needs a proper phase averaging implementation.

\section{Parameter estimation}

In order to infer the formulas describing the accuracy of the estimation of $f_\text{meas}$ and $n_\text{meas}$, we have realized various sets of simulation where we scan the value of the main three parameters $\Nwf$, $\Nmode$ and $n$.

\begin{center}
\begin{figure*}[th]
\includegraphics[width=0.80\textwidth]{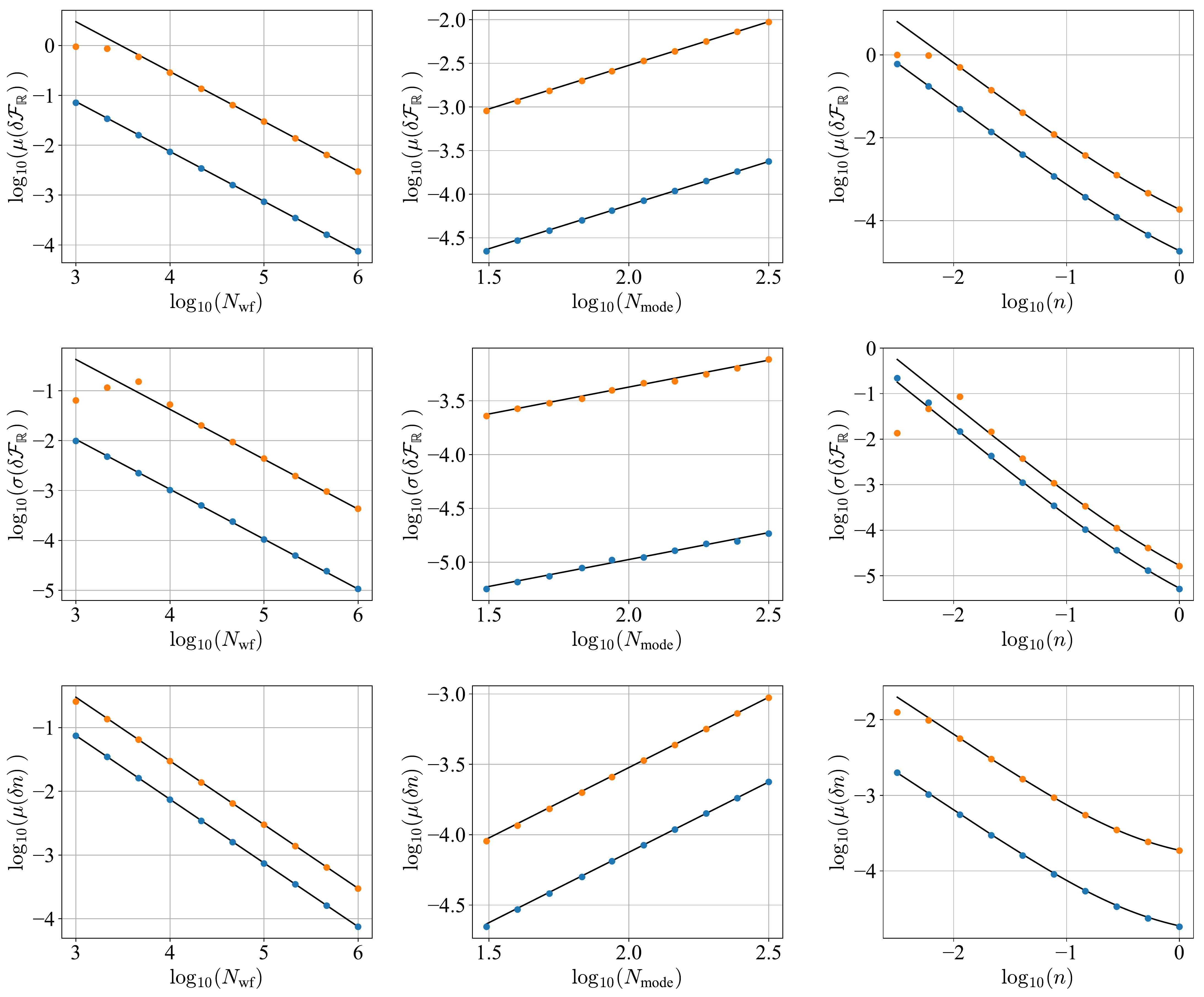}
\caption{Simulation with reconstruction assuming a real-valued temporal mode function. First column: Scans of the parameter $\Nwf$, in blue for the parameters $\Nmode=100$ and $n=1.0$, in orange for $\Nmode=100$ and $n=0.1$.
Second column: Scans of the parameter $\Nmode$, in blue for the parameters $\Nwf=1$M and $n=1.0$, in orange for $\Nwf=1$M and $n=0.1$.
Third column: Scans of the parameter $n$, in blue for the parameters $\Nmode=25$ and $\Nwf=1$M, in orange for $\Nmode=250$ and $\Nwf=1$M.
}
\label{fig:simulation_real}
\end{figure*}
\end{center}

Figure \ref{fig:simulation_real} shows some of the simulation results used to infer the formulas exposed in the main text. The plain black lines are not fits but the corresponding equations given in the main text. For each case we use two sets of parameters to show that the variables are independent. 

\begin{center}
\begin{figure*}[bh]
\includegraphics[width=0.80\textwidth]{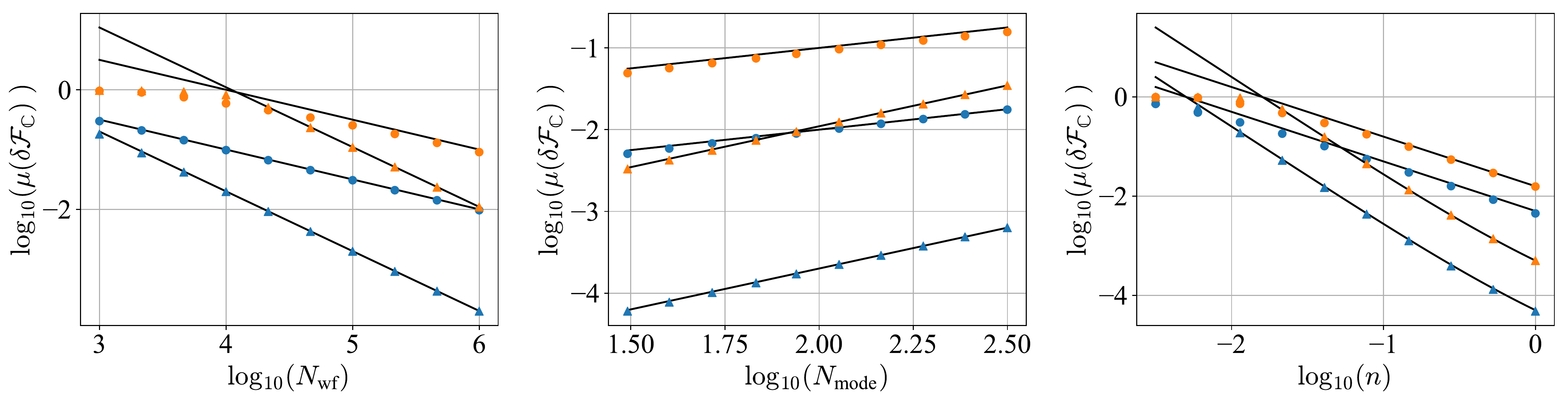}
\caption{Comparison for a complex-valued reconstruction for a real temporal function (circle) and a complex-valued function with balanced photon numbers (triangle). Left graph, scans of the parameter $\Nwf$, in blue for the parameters $\Nmode=100$ and $n=1.0$, in orange for $\Nmode=100$ and $n=0.1$.
Middle graph, scans of the parameter $\Nmode$, in blue for the parameters $\Nwf=1$M and $n=1.0$, in orange for $\Nwf=1$M and $n=0.1$.
Left graph, scans of the parameter $n$, in blue for the parameters $\Nmode=25$ and $\Nwf=1$M, in orange for $\Nmode=250$ and $\Nwf=1$M.
}
\label{fig:simulation_complex}
\end{figure*}
\end{center}

Figure \ref{fig:simulation_complex} corresponds to the simulation in the case of complex-valued function reconstructions for a real function and complex function with balanced eigenvalues (discussion in the main text).

The simulations were performed over the following ranges:
\begin{itemize}
    \item $\Nwf$ over $10^2-10^7$,
    \item $\Nmode$ over $20-500$,
    \item $n$ over $10^{-3}-10^2$.
\end{itemize}
Therefore, we are confident that the inferred formulas are reliable at least over these ranges.

Running simulations beyond those ranges leads to multiple difficulties: Computation time, memory capacity, numerical accuracy and random number generator cycles. However, a large majority of practical implementations are already covered by the investigated parameter ranges.

\onecolumngrid

\twocolumngrid

\end{document}